\documentclass[aps,prc,preprint,groupedaddress]{revtex4-1}

\usepackage{graphicx}
\usepackage{amssymb}
\usepackage{amsmath}
\usepackage{color}

\begin{document}


\title{Nuclear spin dependence of time reversal invariance violating effects in neutron scattering}



\author{Vladimir Gudkov}
\email[]{gudkov@sc.edu}
\affiliation{Department of Physics and Astronomy, University of South Carolina, Columbia, South Carolina 29208, USA}

\author{Hirohiko M. Shimizu}
\email[]{shimizu@phi.phys.nagoya-u.ac.jp}
\affiliation{Department of Physics, Nagoya University, Nagoya 464-8602, Japan}


\date{\today}

\begin{abstract}
The  spin structure of  parity violating and time reversal invariance violating effects in neutron scattering is discussed. The explicit relations between  these effects are presented in terms of  functions  nuclear spins and  neutron partial widths of $p$-wave resonances.
\end{abstract}


\maketitle


\section{Introduction}

Time reversal invariance violating (TRIV)
  effects in nuclear reactions which can be measured in the transmission of polarized neutrons through a polarized target\cite{Kabir:1982tp,Stodolsky:1982tp} attract interest since they   can be  enhanced \cite{Bunakov:1982is,Gudkov:1991qg} by a factor as large as $10^6$.
     Similar  enhancement \cite{Flambaum:1980hg,Sushkov:1982fa}  was already observed for parity violating (PV) effects related  in neutron transmission through nuclear targets. For example, the PV asymmetry in the $0.734 \ {\rm eV}$ $p$-wave resonance in $^{139}{\rm La}$ has been measured to be $(9.56 \pm 0.35)\cdot 10^{-2}$ (see, for example, Ref. \cite{Mitchell2001157} and references therein).
   The recent proposals for searches TRIV in neutron-nucleus scattering (see, for example, Ref. \cite{Bowman:2014fca} and references therein)  demonstrated the existence of a class of experiments that are free from false asymmetries, which can have a discovery potential that is   $10^2$ to $10^4$ more sensitive than current limits due to the enhancement of TRIV effects  for neutron energies corresponding to $p$-wave resonances in the compound nuclear system.
   Also, a search for TRIV in neutron transmission gives the opportunity to use many different   nuclear targets.  This provides assurance  that  possible ``accidental'' cancellation of T-violating effects due to unknown structural factors related to the strong interactions in the particular nucleus can be avoided.  Taking into account that different models of the CP-violation may contribute differently to a particular T/CP-odd observable, which  may have  unknown theoretical uncertainties, TRIV nuclear effects could be considered   complementary   to  electric dipole moment (EDM) measurements, whose status as null tests of T invariance is more widely known.
   It has been shown \cite{Gudkov:1990tb,Gudkov:1991qg} that the ratio of TRIV effects to PV effects  measured at the same $p$-wave resonance has  almost complete cancellation of nuclear reaction effects resulting in a ratio of TRIV to PV matrix elements taken between the same states with opposite parities. This eliminates a bulk of  nuclear   model uncertainties in TRIV effects. \footnote{It should be noted that we do not discuss here the relations between  the ratio of  TRIV to PV matrix elements and the corresponding  ratio of nucleon coupling constants. It was shown \cite{Gudkov:1990tb} that these two ratios have the same order of magnitude with a possible suppression for the  nucleon coupling constants by about a factor of $\sim 3$, which is consistent with the detailed statistical analysis \cite{Flambaum:1993nn,Flambaum:1994kw,Flambaum:1994sd,Flambaum:1997pa} of the correlator of TRIV and PV matrix elements leading to the suppression factor by about of $\sim A^{1/3}$ ( $A$ is the atomic number).} The coefficient of proportionality  between the observables and the corresponding matrix elements involves neutron partial widths with different channel spins, which are unknown and must be determined from independent measurements of angular correlations in neutron induced reactions (see, for example \cite{Sushkov:1985ng}). Since these coefficients have a ``natural'' value of an order of 1, they were mostly ignored in previous studies of TRIV effects in neutron scattering. However, for development of experimental proposals,   it is very important to know the exact spin structure in the relation of PV and TRIV effects to be able to choose  the optimal  target  for TRIV experiments and for the further analysis of the experimental data.

\section{General formalism}

 TRIV and parity violating (PV) effects, related to the T-odd correlation, $\vec{\sigma}\cdot(\hat{\vec{k}}\times\hat{\vec{I}})$ can be observed in the transmission of polarized neutrons through the polarized target, where
$\vec{\sigma}$, $\hat{\vec{I}}$, and $\hat{\vec{k}}$ are unit vectors parallel to the neutron spin, the target nuclear spin and the incident neutron momentum.
 This correlation leads to a P-odd and T-odd
difference  between the total neutron cross sections \cite{Stodolsky:1982tp} $\Delta\sigma_{\rm PT}$  for $\vec{\sigma}$
parallel and anti-parallel to $\hat{\vec{k}}\times\hat{\vec{I}}$
\begin{equation}
\label{delT}
\Delta\sigma_{\rm PT}=\frac{4\pi}{k}{\rm Im}(f_{\uparrow}-f_{\downarrow}).
\end{equation}
Here, $f_{\uparrow,\downarrow}$ are the zero-angle scattering amplitudes for neutrons polarized
parallel and anti-parallel to the $\vec{k}\times{\vec{I}}$ axis, respectively.
These TRIV effects  can be  enhanced \cite{Bunakov:1982is} by a factor as large as $10^6$. Similar  enhancement was already observed for PV effects related to  $(\vec{\sigma}\cdot\hat{\vec{k}})$ correlation in neutron transmission through unpolarized nuclear targets.
  This PV and
TRI-conserving difference of total cross sections $\Delta\sigma_{\rm P}$  can be written in terms of
differences of zero angle elastic scattering amplitudes with negative  and
positive  neutron helicities as:
\begin{equation}\label{delP}
 \Delta \sigma_{\rm P} = {{4\pi}\over k}{\rm Im}(f_- - f_+).
\end{equation}

To obtain the explicit spin dependent coefficients for these two observables  we consider the reaction matrix $\hat{T}$, which is related to the scattering matrix $\hat{\mathbb{S}}$ and the matrix $\hat{R}$ as
\begin{equation}\label{matr}
2\pi i \hat{T}=\hat{1}-\hat{\mathbb{S}}=\hat{R}.
\end{equation}

It is convenient to relate this matrix to the matrix $\hat{R}$ in the integral of motion representation of the $S$-matrix \cite{Baldin:1961}
\begin{equation}\label{Smat}
\left\langle S^{\prime} l^{\prime} \alpha^{\prime} \middle\vert \mathbb{S}^J \middle\vert S l \alpha \right\rangle
\delta_{JJ^{\prime}}\delta_{MM^{\prime}}\delta(E^{\prime} -E) ,
\end{equation}
where $J$ and $M$ are the total spin and its projection, $S$ is the channel spin, $l$ is the orbital momentum, and $\alpha$ represents the other internal quantum numbers.
Taking into account that the spin channel is a sum of the neutron spin $\vec{s}$ and nucleus spin $\vec{I}$
\begin{equation}\label{scheam1}
\vec{S}=\vec{s}+\vec{I},
\end{equation}
 the total angular momentum is
\begin{equation}\label{schem2}
 \vec{J}=\vec{l}+\vec{S}.
\end{equation}
Then by choosing quantization axis along axis $z$,   one can write $T$-matrix elements for elastic forward scattering  as
\begin{eqnarray}\label{genMat}
 \nonumber
  2\pi i
  \left\langle \vec{k}\mu \middle\vert T \middle\vert \vec{k}\mu \right\rangle
  &=& \sum_{JM_Jlml^{\prime} m^{\prime} S m_s S^{\prime} m^{\prime}_s}Y_{l^{\prime} m^{\prime}}(\theta ,\phi)
  	\left\langle s \mu I M \middle\vert S^{\prime} m^{\prime}_s \right\rangle
	\left\langle l^{\prime} m^{\prime} S^{\prime} m^{\prime}_s  \middle\vert JM_J \right\rangle\\
  &\times&
  	\left\langle S^{\prime} l^{\prime} \alpha^{\prime} \middle\vert R^J \middle\vert S l \alpha\right\rangle
	\left\langle JM_J \middle\vert l m S m_s  \right\rangle
	\left\langle S m_s \middle\vert s \mu I M \right\rangle
	Y^*_{lm} (\theta ,\phi) ,
 \end{eqnarray}
where angles $(\theta ,\phi)$ describe the direction of the neutron momentum $\vec{k}$, and $\mu$ is the projection of the neutron spin along the axis of quantization.

Using the relation for a reaction amplitude $\hat{f}$:  $\hat{f}=-\pi(k_{\rm i} k_{\rm f})^{-1/2}\hat{T}$, where $k_{\rm i,f}$ are values of initial and final momenta, respectively,
we can write the elastic scattering amplitude as

\begin{eqnarray}\label{genAmp}
 \nonumber
 f  & =\frac{i}{2k}& \sum_{JM_Jll^{\prime}  S m_s S^{\prime} m^{\prime}_s}Y_{L m_L}(\theta ,\phi)
  	\left\langle s \mu I M \middle\vert S^{\prime} m^{\prime}_s \right\rangle
	\left\langle S m_s \middle\vert s \mu I M \right\rangle\\
  &\times&
   \nonumber
  	\left\langle S^{\prime} l^{\prime} \alpha^{\prime} \middle\vert R^J \middle\vert S l \alpha\right\rangle
	(-1)^{J+S^{\prime}+l^{\prime}+l}(2J+1)\sqrt{\frac{(2l+1)(2l^{\prime}+1)}{4\pi (2S+1)}} \\
	 &\times&
\left\langle l 0 l^{\prime} 0 \middle\vert L 0 \right\rangle
\left\langle L m_L S^{\prime} m^{\prime}_s \middle\vert S m_s \right\rangle
 \begin{Bmatrix}
   l^{\prime} & l & L \\
   S & S^{\prime} & J
  \end{Bmatrix}
 \end{eqnarray}

For the further calculations we   assume that  the direction of   the target polarization  $\vec{I}$ is along axis $z$ , and the direction of the momentum $\vec{k}$ is along the axis $y$. Then, for low energy neutrons ($l, l^{\prime}\leq 1$) we can obtain from Eq.(\ref{genAmp}) the differences of the amplitudes related to TRIV correlation $(\vec{\sigma}\cdot [\hat{\vec{k}}\times\hat{\vec{I}}])$ and P-odd correlation $(\vec{\sigma}\cdot\hat{\vec{k}})$ as

\begin{eqnarray}\label{Tamp}
  & & \Delta f_{\rm PT}=f_{\uparrow}-f_{\downarrow} =\\
  \nonumber
      & &  M\frac{\sqrt{3}I }{8 \pi  k \sqrt{2 I+1}}\left( \frac{{\left\langle (I-1/2), 0 \middle\vert R^{I-1/2} \middle\vert (I+1/2),1 \right\rangle}-{\left\langle (I+1/2), 1 \middle\vert R^{I-1/2} \middle\vert (I-1/2),0 \right\rangle}}{\sqrt{I+1}} \right.\\
       \nonumber
   &+& \left. \frac{{\left\langle (I+1/2), 0 \middle\vert R^{I+1/2} \middle\vert (I-1/2),1 \right\rangle}-{\left\langle (I-1/2), 1 \middle\vert R^{I+1/2} \middle\vert (I+1/2),0 \right\rangle}}{\sqrt{I}} \right),
\end{eqnarray}

and

\begin{eqnarray}\label{Pamp}
 & &\Delta f_{\rm P} =f_+-f_-=-\frac{i }{4 \pi  \sqrt{6 I+3} k} \\
  \nonumber
 &\times &\left\{ I \sqrt{2 I-1}
   \left[ {\left\langle (I-1/2), 0 \middle\vert R^{I-1/2} \middle\vert (I-1/2),1 \right\rangle}) +{\left\langle (I-1/2), 1 \middle\vert R^{I-1/2} \middle\vert (I-1/2),0 \right\rangle}\right]\right. \\
    \nonumber
   &+&2 \sqrt{I} (I+1)
   \left[ {\left\langle (I+1/2), 0 \middle\vert R^{I+1/2} \middle\vert (I-1/2),1 \right\rangle}) +{\left\langle (I-1/2), 1 \middle\vert R^{I+1/2} \middle\vert (I+1/2),0 \right\rangle}\right] \\
    \nonumber
   &-&2 I \sqrt{I+1}
   \left[ {\left\langle (I-1/2), 0 \middle\vert R^{I-1/2} \middle\vert (I-1+/2),1 \right\rangle}) +{\left\langle (I+1/2), 1 \middle\vert R^{I-1/2} \middle\vert (I-1/2),0 \right\rangle}\right]\\
    \nonumber
   &-&(I+1) \sqrt{2 I+3}\left.
   \left[ {\left\langle (I+1/2), 0 \middle\vert R^{I+1/2} \middle\vert (I+1/2),1 \right\rangle}) +{\left\langle (I+1/2), 1 \middle\vert R^{I+1/2} \middle\vert (I+1/2),0 \right\rangle}\right]\right\} .
 \end{eqnarray}

Both TRIV and PV  amplitudes can be calculated using the distorted wave Born
approximation to first order in the parity and time reversal
violating interactions (see, for example, Ref.\cite{Bunakov:1982is}).
Then for slow neutrons the PV and TRIV  matrix elements in above expressions   can be written in the Breit-Wigner resonance approximation with one $s$ resonance and one $p$ resonance as \cite{Bunakov:1982is}
\begin{equation}\label{BWamp}
\left\langle S^{\prime} l^{\prime} \middle\vert R^{J} \middle\vert S l \right\rangle =
 \frac
{\sqrt{\Gamma^{\rm n}_{l^{\prime}}(S^{\prime})}(-iv+w)\sqrt{\Gamma^{\rm n}_{l}(S)}}
{(E-E_l+i\Gamma_l/2)(E-E_{l^{\prime}}+i\Gamma_{l^{\prime}}/2)}
e^{i(\delta_{l^{\prime}}(S^{\prime})+\delta_{l}(S))}
\end{equation}
where $E_l$, $\Gamma_l$, and $\Gamma^{\rm n}_{l}(S)$ are the energy, the total width, and the partial neutron width of the $l$-th nuclear compound resonance, $E$ is the neutron energy, and $\delta_{l}$ is the potential scattering phase shift, $l\neq l^{\prime}$, and $v$ and $w$ are PV and TRIV nuclear matrix elements between $s$-wave and $p$-wave compound states, correspondingly:
$$< \varphi_{\rm s}|V_{\rm P}+W_{\rm PT}|\varphi_{\rm p} > =< \varphi_{\rm s}|V_{\rm P}|\varphi_{\rm p} >+< \varphi_{\rm s}|W_{\rm PT}|\varphi_{\rm p} >= -v-iw .$$
Here $\varphi_{\rm s}$ and $\varphi_{\rm p}$ are compound states wave functions, and $V_{\rm P}$ and $W_{\rm PT}$ are PV and TRIV interactions nuclear operators. Therefore (see, for example, Refs.  \cite{Bunakov:1982is,Gudkov:1990tb,Gudkov:1991qg,Bowman:2014fca}), at each resonance the ration of TRIV and PV effects is related to the ration of TRIV and PV nuclear matrix elements multiplied by a spin-dependent factor $\kappa$
\begin{equation}\label{kappa}
\frac{\Delta\sigma_{\rm PT}}{\Delta\sigma_{\rm P}} = \kappa \frac{w}{v}.
\end{equation}
The important point is that the factor $\kappa$ includes amplitudes of the partial neutron widths which depend on spin channels. To see this dependence let us consider the ratio of TRIV and PV effects at resonances with a total spin $J=I+1/2$ and $J=I-1/2$ separately, assuming 100\% polarization of the target for TRIV effect ($M=I$) and unpolarized target for PV effect. Then, using Eqs. (\ref{Tamp}), (\ref{Pamp}), and (\ref{BWamp}), we obtain
\begin{eqnarray}\label{kappaPdef}
& &  \kappa (J=I+1/2) = \left(\frac{\Delta\sigma_{\rm PT}}{\Delta\sigma_{\rm P}} \right)_{J=I+1/2} \\
  \nonumber
  &=& \left( \frac{3\sqrt{I}(2I+1)}{2(I+1)}\right) \frac{\sqrt{\Gamma^{\rm n}_{\rm p}(I-1/2)}}{[ \sqrt{2I+3}\sqrt{\Gamma^{\rm n}_{\rm p}(I+1/2)} -2\sqrt{I}\sqrt{\Gamma^{\rm n}_{\rm p}(I-1/2)}]},
\end{eqnarray}
and
\begin{eqnarray}\label{kappaMdef}
& &  \kappa (J=I-1/2) = \left(\frac{\Delta\sigma_{\rm PT}}{\Delta\sigma_{\rm P}} \right)_{J=I-1/2} \\
  \nonumber
  &=& \left( \frac{3(2I+1)}{2\sqrt{I+1}}\right) \frac{\sqrt{\Gamma^{\rm n}_{\rm p}(I+1/2)}}{[ 2\sqrt{I+1}\sqrt{\Gamma^{\rm n}_{\rm p}(I+1/2)}-\sqrt{2I-1}\sqrt{\Gamma^{\rm n}_{\rm p}(I-1/2)}]}.
\end{eqnarray}

For the further analysis it is convenient to use a ratio of $[{\Gamma^n_{p}(I+1/2)}/{\Gamma^n_{p}(I-1/2)}]^{1/2}\equiv \gamma$ and relative fractions of the amplitudes of the neutron decay widths
\begin{eqnarray}\label{xSyS}
  x_S &=& [{\Gamma^{\rm n}_{\rm p}(I-1/2)}/{\Gamma^{\rm n}_{\rm p}}])^{1/2}\\
  y_S &=& [{\Gamma^{\rm n}_{\rm p}(I+1/2)}/{\Gamma^{\rm n}_{\rm p}}])^{1/2},
\end{eqnarray}
where ${\Gamma^{\rm n}_{\rm p}}={\Gamma^{\rm n}_{\rm p}(I+1/2)}+{\Gamma^{\rm n}_{\rm p}(I-1/2)}$ is a total  $p$-wave neutron width. Then, Eqs.(\ref{kappaPdef}) and (\ref{kappaMdef}) can be written as\footnote{These expressions are different from the presented ones in previous papers \cite{Gudkov:1990tb,Gudkov:1991qg,Bowman:2014fca}, where a misprint from Ref. \cite{Gudkov:1990tb} was propagated without been noticed.}
\begin{eqnarray}
 \nonumber
  \kappa (I+1/2) &=& \left( \frac{3\sqrt{I}(2I+1)}{2(I+1)}\right) \frac{1}{[\gamma \sqrt{2I+3} -2\sqrt{I}]} \\
  \kappa (I-1/2) &=& \left( \frac{3(2I+1)}{2\sqrt{I+1}}\right)\frac{1}{[2\sqrt{I+1}-\sqrt{2I-1}/\gamma ]}
\end{eqnarray}

Unfortunately, the parameters $\gamma$, $x_S$, and $y_S$ are unknown,  and must be obtain from additional experiments, for example, from the measurements of different angular correlations in $(n,\gamma )$ reactions. The complete analysis of angular correlations in $(n,\gamma )$ reactions for low energy neutrons has been done in Ref. \cite{Sushkov:1985ng}, where
 a different spin coupling scheme was used:  $\vec{J}=(\vec{s}+\vec{l})+\vec{I}$.  The relations between these two spin coupling schemes are given in Appendix. In the coupling scheme used in Ref. \cite{Sushkov:1985ng} one can define the corresponding fractions of the amplitudes of neutron widths as
\begin{eqnarray}\label{xy}
  x &=& ({\Gamma^{\rm n}_{{\rm p}1/2}}/{\Gamma^{\rm n}_{\rm p}})^{1/2}\\
  y &=& ({\Gamma^{\rm n}_{{\rm p}3/2}}/{\Gamma^{\rm n}_{\rm p}})^{1/2},
\end{eqnarray}
where $\Gamma^{\rm n}_{{\rm p}j}$ correspond to $j=1/2$ and $j=3/2$  with $\vec{j}=\vec{s}+\vec{l}$, and ${\Gamma^{\rm n}_{\rm p}}={\Gamma^{\rm n}_{{\rm p}1/2}}+{\Gamma^{\rm n}_{{\rm p}3/2}}$.

Then, taking into account that
for $J=I+1/2$,
\begin{equation}\label{Jp}
[\Gamma^{\rm n}_{{\rm p}1/2}]^{1/2}=\frac{1}{\sqrt{3(2I+1)}}( -2\sqrt{I}[\Gamma^{\rm n}_{\rm p}(I-1/2)]^{1/2}+\sqrt{2I+3}[\Gamma^{\rm n}_{\rm p}(I+1/2)]^{1/2}),
\end{equation}
and for $J=I-1/2$,
\begin{equation}\label{Jm}
[\Gamma^{\rm n}_{{\rm p}1/2}]^{1/2}=\frac{1}{\sqrt{3(2I+1)}}( -\sqrt{2I-1}[\Gamma^{\rm n}_{\rm p}(I-1/2)]^{1/2}+2\sqrt{I+1}[\Gamma^{\rm n}_{\rm p}(I+1/2)]^{1/2}),
\end{equation}
we can rewrite  Eqs.(\ref{kappaPdef}) and (\ref{kappaMdef}) as
\begin{eqnarray}
 \nonumber
  \kappa (I+1/2) &=& \left( \frac{\sqrt{3I(2I+1)}}{2(I+1)}\right)\sqrt{\frac{\Gamma^{\rm n}_{\rm p}(I-1/2)}{\Gamma^{\rm n}_{{\rm p}1/2}}}   \\
  \kappa (I-1/2) &=& \left( \frac{\sqrt{3(2I+1)}}{2\sqrt{(I+1)}}\right)\sqrt{\frac{\Gamma^{\rm n}_{\rm p}(I+1/2)}{\Gamma^{\rm n}_{{\rm p}1/2}}},
\end{eqnarray}
or as
\begin{eqnarray}
 \nonumber \label{kappaxy}
  \kappa (I+1/2) &=& \left( \frac{\sqrt{I}}{2(I+1)}\right)\frac{[-2\sqrt{I}x+\sqrt{2I+3}y]}{x}   \\
  \kappa (I-1/2) &=& \left( \frac{1}{2\sqrt{(I+1)}}\right)\frac{[2\sqrt{I+1}x+\sqrt{2I-1}y]}{x}.
\end{eqnarray}

Since the parameter $\kappa$ is a coefficient of  proportionality between the observed value of PV effect and the unknown value of TRIV effect, it is very important to know $\kappa$ for the estimation of the accuracy of TRIV effects in the proposed experiments. The lager value of   the $\kappa$ leads to better statistics in the measurements of TRIV effects.
The obtained expressions for the $\kappa$ parameter gives us the opportunity to calculate explicitly the ratio of TRIV and PV effects at each $p$-wave nuclear compound resonance in terms of the amplitudes of neutron widths, or in terms of two sets  of parameters: $x$ and $y$ or $x_s$ and $y_s$, which satisfy the constrains $x^2+y^2=1$ and $x_s^2+y_s^2=1$, correspondingly.  The  values of amplitudes of neutron widths are randomly fluctuated on different $p$-wave compound  resonances for the same nuclei, which may lead to  different  sensitivities of  TRIV effects for different resonances at the same nuclear target.
 However, there is one exceptional case   for  $J=0$ $p$-wave resonances for nuclei with spin $1/2$ where the parameter $\kappa $ does not depend on neutron partial widths at all, and, as a consequence, can be calculated exactly as $\kappa =1 $ [see eq.(\ref{kappaxy})]. Another interesting observation from Eq.(\ref{kappaxy}) is related to the fact that PV effects are proportional to the parameter $x$ [or to $\Gamma^{\rm n}_{{\rm p}1/2}$], while TRIV effects to the linear superposition of $x$ and $y$ parameters (or to  $\Gamma^{\rm n}_{\rm p}(I\pm1/2)$). Therefore, TRIV effects can be observed even on p-wave resonances which do not show visible PV effects.

 In spite of the fact that total $p$-wave  neutron widths are measured for many p-wave resonances with a good accuracy \cite{Mitchell2001157,Mughabghab}, the parameters $x$ and $y$ (or $x_s$ and $y_s$) currently are practically unknown. They cannot be calculated due to a complex structure of compound resonances, therefore,   they can be obtained only  from experiments which are sensitive to the values and signs of amplitudes of neutron widths. For example, as  was mentioned above, they can be obtain from  angular distributions of $\gamma$ rays in resonance neutron radiative capture reactions or by measuring of pseudomagnetic neutron spin rotation in the vicinity of $p$-wave resonances \cite{Gudkov:2017sye}.

\section{Discussions and Conclusions}

The obtained  expressions of spin-dependent factors for TRIV and PV effects give us the opportunity to clarify a number of issues for the proposed search for TRIV effects in transmission of polarized neutron through polarized targets.
 We can see that the parameter $\kappa$ has a ``natural'' value of an order of unity with the possible fluctuations on different resonances due to unknown spin-dependent  partial neutron width amplitudes of $p$-wave resonances.
 In some cases this can lead to a quite significant enhancement or suppression of TRIV effects at different p-wave resonances. Therefore, it is very important to measure these neutron partial widths for the choice of the optimal target. Also, one can see that, depending on the  values of the neutron partial widths, resonances with the largest PV effects may not  have the largest TRIV effects, and otherwise one can observe reasonably large TRIV effects at $p$-wave resonances with a ``regular'' or small values of PV effects. This suggest that all $p$-wave resonances with  observed PV effects should  be tested   for the possible TRIV experiments by measuring their partial neutron widths.

 The general expressions in Eqs.(\ref{Tamp}) and (\ref{delT}) show that TRIV effects are proportional to the projection $M$ of nuclear target spin on the axis of quantization. Therefore,   for the case of a  target with a superposition of different polarization modes  (vector, tensor, etc.), only a vector part of the polarization leads to TRIV effects, and  the value of TRIV effect is modified by a factor
\begin{equation}\label{polar}
P =  \frac{1}{I}\sum_{M} w(M)M,
\end{equation}
where $w(M)$ is a weight of a population for each spin projection quantum number $M$, or a weight in the density operator used for the description of the general polarization in terms of the density polarization matrix.

Finally, we would like to note again that the expressions for the parameter $\kappa$ obtained in this paper are different from the presented  in a number of publications (see, for example Refs. \cite{Gudkov:1990tb,Gudkov:1991qg,Bowman:2014fca}) due to   misprints in Eq.(3) of Ref.  \cite{Gudkov:1990tb} which were propagated to different publications  without being noticed. Fortunately, the misprinted value of the $\kappa$  does not change the results reported in these publications since for all estimates it was assumed that $\kappa$ is unknown parameter of an order of unity. However,   the knowledge of the exact value of  $\kappa$ is required for the choice of the target for the experiment, and for the analysis of experimental data. In this relation, the first measurement of   $\kappa$,   $x$, and $y$ parameters has been done recently \cite{Okudaira:2017dun} for $0.74\; eV$ $p$-wave compound nuclear resonance of $^{139}$La by measuring angular distribution of individual $\gamma$ rays in neutron-induced radiative capture.

\begin{acknowledgments}
This material is based upon work supported by the U.S. Department of Energy Office of Science, Office of Nuclear Physics program under Award No. DE-SC0015882.
\end{acknowledgments}

\appendix

\section{Relations between different spin-coupling schemes}

The relation between two spin coupling schemes $\vec{J}=\vec{l}+(\vec{I}+\vec{s})$ and $\vec{J}=(\vec{s}+\vec{l})+\vec{I}$ is given by
\begin{equation}\label{spincoupl}
\left\langle
	\left(\left(l,\frac12\right)j,I\right)J
\middle\vert
	\left(l,\left(\frac12,I\right)S\right)J
\right\rangle
	=(-1)^{l+1/2+I+J}\sqrt{(2j+1)(2S+1)}\left\{
		\begin{array}{ccc}
			l & \frac12 & j \\
			I & J & S
		\end{array}
	\right\},
\end{equation}
where $\vec{S}=\vec{I}+\vec{s}$ and $\vec{j} =\vec{s}+\vec{l}$.

Now, defining
\begin{eqnarray}
  x &=& \left\vert j=\frac12 \right\rangle \\
  \nonumber
  y &=& \left\vert j=\frac32 \right\rangle \\
  \nonumber
  x_s &=& \left\vert S=I-\frac12 \right\rangle \\
  \nonumber
  y_s &=& \left\vert S=I+\frac12 \right\rangle ,
\end{eqnarray}
one can write for $l=1$
\begin{eqnarray}
x_s &=& (-1)^{3/2+I+J}\sqrt{4I}\left\{
	\begin{array}{ccc}
		1 & \frac{1}{2}& \frac{1}{2} \\
		I & J & I-\frac12
	\end{array}
	\right\} x
	+(-1)^{3/2+I+J}\sqrt{8I}\left\{
	\begin{array}{ccc}
		1 & \frac{1}{2} & \frac{3}{2} \\
		I & J & I-\frac12
	\end{array}
	\right\}y\\
 	\nonumber
y_s &=& (-1)^{3/2+I+J}\sqrt{4(I+1)}\left\{
	\begin{array}{ccc}
		1 & \frac{1}{2} & \frac{1}{2} \\
		I & J & I+\frac12
	\end{array}
	\right\} x
	+(-1)^{3/2+I+J}\sqrt{8(I+1)}\left\{
	\begin{array}{ccc}
		1 & \frac{1}{2} & \frac{3}{2} \\
		I & J & I+\frac12
	\end{array}
	\right\}y .
\end{eqnarray}


\bibliography{TViolation,ParityViolation}

\end{document}